\documentclass[fleqn,usenatbib]{mnras}

\usepackage{epstopdf}
\usepackage{newtxtext,newtxmath}
\usepackage{txfonts}
\usepackage[utf8]{inputenc}
\usepackage[T1]{fontenc}

\usepackage[T1]{fontenc}
\usepackage{ae,aecompl}
\usepackage{multicol}

\usepackage{graphicx}	% Including figure files
\usepackage{aas_macros} % typeset journal abbreviations correctly.
\usepackage{amsmath}	% Advanced maths commands
\usepackage{amssymb}	% Extra maths symbols

\usepackage[usenames]{color}

\newcommand{\red}{} %\textcolor{red} 

\title[The pulsating binary, TT\,Hor]{A window into $\delta$~Sct stellar interiors: Understanding the eclipsing binary system TT~Hor }

\author[M. Streamer et al.]{
Margaret Streamer$^{1}$\thanks{E-mail: margaret.streamer@anu.edu.au (MS)},
Michael J. Ireland$^{1}$,
Simon J. Murphy$^{2,3}$
and Joao Bento$^{1}$
\\
$^{1}$ Research School of Astronomy and Astrophysics, Australian National University, Canberra, ACT, 2611, Australia  \\
$^{2}$ Sydney Institute for Astronomy (SIfA), School of Physics, The University of Sydney, NSW 2006, Australia \\
$^{3}$ Stellar Astrophysics Centre, Department of Physics and Astronomy, Aarhus University, DK-8000 Aarhus C, Denmark\\
}

\date{Accepted XXX. Received YYY; in original form ZZZ}

\pubyear{2017}

\begin{document}
\label{firstpage}
\pagerange{\pageref{firstpage}--\pageref{lastpage}}
\maketitle

% Abstract of the paper
\begin{abstract}
The semi-detached eclipsing binary system, TT\,Hor, has a $\delta$~Sct primary component (accretor) accreting mass from the secondary star (donor).  We fit an eclipsing binary  model from V, B and I photometry combined with spectroscopy using {\sc phoebe}.  Radial velocity variations of the center of mass of TT\,Hor AB over 2 years suggest the presence of a wide companion, consistent with a Kozai-Lidov resonance formation process for TT\,Hor AB.   Evolutionary models computed with \textsc{mesa} give the initial mass of the donor as $\approx$1.6 $M_{\sun}$ and that of the accretor as $\approx$1.3\,$M_{\sun}$.  The initial binary orbit has a similar initial separation to the currently observed separation of 11.4 $R_{\sun}$.  Mass transfer commences at an age of 2.5\,Gyr when the donor is a subgiant.  We model the accretor as a tidally-locked, 2.2 $\pm$ 0.2 -$M_{\sun}$  $\delta$~Sct pulsator which has accreted $\approx$0.9\,$M_{\sun}$ of slightly He-enriched material (mean Delta Y <0.01) from the donor over the last \red{90}\,Myr. The best fit from all measured parameters and evolutionary state is for a system metallicity of [M/H] is \red{0.15.} A pulsation model of the primary gives a self-consistent set of modes. Our observed oscillation frequencies match to within \red{0.3}$\%$ and the system parameters within uncertainties.  However, we cannot claim that our identified modes are definitive, and suggest follow-up time-series spectroscopy at high resolution in order to verify our identified modes.  With the higher SNR and continuous observations with TESS, more reliable mode identification due to frequency and amplitude changes during the eclipse is likely.
\end{abstract}

% Select between one and six entries from the list of approved keywords.
% Don't make up new ones.

%%%%%%%%%%%%%%%%%%%%%%%%%%%%%%%%%%%%%%%%%%%%%%%%%%

%%%%%%%%%%%%%%%%% BODY OF PAPER %%%%%%%%%%%%%%%%%%

\section{Introduction}

A-stars span the transitional area of energy transport and can exhibit both a tiny convective core and a tiny convective envelope overlying a radiative envelope. They are often modelled, for example, by assuming homogeneous heavy element (CNO) abundances and neglecting rotation \citep{noels_-type_2004}.  However, many A-stars are complex and rotate rapidly (>$100$\,km\,s$^{-1}$)  \citep{royer_rotational_2007} and the ongoing challenge is to incorporate these complexities into structural models and stellar evolution codes.

A-stars that exhibit $\delta$~Sct (pressure-mode) pulsations present an opportunity to delve into the internal structure of these stars and improve models of their structure and evolution.  In recent years, asteroseismology has given exquisite information about the internal structure of our Sun and other solar-like oscillators (see \citealt{bedding_solar-like_2014}, for a perspective). The evolutionary stages of red giant stars have also been investigated successfully using asteroseismology (e.g., \citealt{kallinger_evolutionary_2012} and a recent review by \citealt{hekker_giant_2017}). 

Conversely, accurate asteroseismology of $\delta$~Sct stars remains elusive because of the difficulty of mode identification from the complex oscillation spectra.  Models and observational data for $\delta$~Sct stars do not always coincide. For example, 4\,CVn has been studied extensively for over 40 years without reaching a satisfactory model (e.g., \citep{schmid_discovery_2014,breger_nonradial_2017}). Rotation, convection, metallicity and magnetism all affect the type of $\delta$~Sct pulsations observed.  Of these, rotational splitting of the non-radial modes, made even worse in fast rotating stars, has a major impact on mode identification \citep{aerts_asteroseismology_2010}.  However, to successfully interpret the internal structure of the $\delta$~Sct star from observed pulsation frequencies, the effect of rotation must be described correctly in the first place.

The number of observable frequencies detected from space missions ($\it Kepler$ and CoRoT) have increased multi-fold with several theoretical considerations relating stellar properties to asteroseismology now published.  One outstanding example is the analysis by \citet{yu_asteroseismology_2018} of 16094 oscillating red giants observed by $\it Kepler$. This catalogue presents asteroseismic parameters from which mass, radius and surface gravity are derived for red giants.  However, far from being enlightening, the plethora of data for $\delta$~Sct stars has presented challenges to their theoretical interpretation.  \citet{garcia_hernandez_observational_2015}, 
found a scaling relation between the large frequency separation, $\Delta$$\upsilon$, and the mean density of $\delta$~Sct stars which confirmed the theoretical prediction of \citet{suarez_measuring_2014}.   \citet{garcia_hernandez_precise_2017} have also calculated log\textit{g} values by correlation  with mean density, independent of the rotation rate.  These latter studies on scaling relations are valuable for the interpretation of the asteroseismology of $\delta$~Sct stars, but for now, we must resort to eclipsing binary systems for the fundamental parameters of these stars.

Clearly, the more knowledge we have about the structure of a $\delta$~Sct star, the better predictions from asteroseismic models will be.  Eclipsing binary systems are ideal to constrain stellar parameters as the absolute masses and radii of the components can be determined with confidence using a combination of photometry and spectroscopy. If one of the components is pulsating then the system is perfect for probing stellar evolution theory and the internal structure leading to pulsations. If tidal locking can be assumed, then the rotational period of the pulsator is also known. With a $\delta$~Sct star that has well-characterised fundamental properties from a binary system and identified pulsation frequencies, more reliable modelling of pulsation modes will be possible.  From this basis, models for single isolated $\delta$~Sct stars can be validated. 

$\delta$~Sct stars exhibit multiple periods from 0.5 to 8hrs with amplitudes from a few $\mu$mag to more than 300 mmag \citep{rodriguez_revised_2000}. They typically pulsate in non-radial, p modes.  Using data from Kepler observations, \citet{grigahcene_hybrid_2010} were the first to identify numerous hybrid  $\delta$~Sct/$\gamma$~Dor Dor pulsators.  Later, \citet{balona_pulsation_2015} suggested that lower frequencies characteristic of $\gamma$~Dor pulsations (non-radial, g modes with periods from one to 5\,d) are present in all $\delta$~Sct stars. However, though rare, pure $\delta$~Sct with no $\gamma$~Dor pulsations have been found \citep{bowman_characterising_2018}.   

\citet{liakos_catalogue_2017} have catalogued 199 cases of eclipsing binary systems with at least one component being a $\delta$~Sct star.  Surprisingly few of these systems have been characterised to any degree.  Some notable exceptions are KIC 3858884, a system in a highly eccentric orbit with a period of 26 days \citep{maceroni_kic_2014} and KIC\,9851944, with a short orbital period of 2.16 days \citep{guo_kepler_2016}.  Both these binary systems have stars of similar masses ($\approx$1.8 $M_{\sun}$) which exhibit both g- and p-mode pulsations.  Additionally, \citet{murphy_finding_2018}  have published a catalogue of 314 $\delta$~Sct pulsators in non-eclipsing binaries. This catalogue triples the number of known intermediate-mass binaries with full orbital parameters.  

Sixty systems in the Liakos catalogue were reliably described as semi-detached, Algol-type systems in which mass transfer occurs.  In these systems, the donor star is the more massive of the pair initially, evolves fastest and then transfers mass to its less massive partner.  Typically, we observe the system when the donor has evolved to the red giant phase, and lost significant mass. The accretor has gained mass, is hotter, and of larger radius than initially. 

In this paper, we consider TT\,Hor: a low-mass Algol binary which exhibits $\delta$~Sct pulsations.  It was first reported as a variable star by  \citet{strohmeier_elements_1967} but the existence of $\delta$~Sct pulsations was not identified until 2013 \citep{moriarty_discovery_2013}.  The system offers the opportunity to probe the structure and evolution of an A-type star undergoing mass transfer.  

We have fully characterised the system by combining time-series photometry with spectroscopic observations and we have analysed the pulsations.  We compare the evolution of the system and that of a single star of similar mass to the pulsator and compute the expected pulsations in both scenarios.  We are able to match the computed frequencies to the observed ones with remarkable success.

\section{OBSERVATIONS AND DATA REDUCTION}

\subsection{Photometry}
Time-series photometry was performed with a 350-mm Meade Schmidt-Cassegrain telescope situated in Murrumbateman, NSW, Australia, IAU observatory code EO7.  The telescope was equipped with either an SBIG ST8XME  or SBIG STT1603ME CCD camera  with Johnson B and V filters, and a Cousins I filter. 

{\sc dimension\,{\small 4}} (Thinking Man Software, 1992 -- 2014, \url{http://www.thinkman.com/dimension4/}) was used to synchronize the computer's clock to UTC.  A fast cadence was used to ensure good coverage of the pulsations and for accurate determination of the eclipse times of minima.  Typical exposure times were of 60 to 120\,s with delays between each exposure (for image download) of about 30\,s.  TT\,Hor was  observed during secondary eclipses and out of eclipse for as long as possible in any one night to maximise the time span available for Fourier analysis of the pulsations.  Data were collected in 2013, 2014 and 2016 and are available on the AAVSO website (\url{https://www.aavso.org/}).

TT\,Hor (RA: 03 27 04.388, Dec: $-$45 52 56.58) is in a sparse field and scrutiny of a high resolution image from the Digitized Sky Survey catalogue (\url{https://archive.stsci.edu/cgi-bin/dss_form/}) and 2MASS \citep{skrutskie_two_2006}  showed no background contaminating stars to affect the photometry.

The data were reduced with aperture photometry using {\sc MaxIm DL}.  GSC\,8059\,0284 was used as the comparison star: V = 10.711 $\pm$ 0.059\,mag, B$-$V = 0.54\,mag. GSC\,8059\,0309 was used as a check star: V = 11.997, B$-$V = 0.68\,mag.  Magnitudes are taken from the AAVSO Photometric All-Sky Survey (APASS).  The APASS \textit{i}' magnitude data were converted to Cousins I filter magnitudes using the conversions from Munari (2012, Henden, private communication).  The data were then transformed using standard stars from Landolt fields \citep{landolt_ubvri_2007}.

TT\,Hor was not included in Gaia data release 1 and so there are no parallax measurements to determine accurate distances and the absolute magnitude of the system.  TT\,Hor is outside the Galactic plane at a latitude of $-54^{\circ}$ where interstellar reddening is minimal and no more than the quoted errors for the comparison  and standard stars.  Therefore the colour magnitudes obtained were not corrected for any interstellar reddening.

Data were also collected in 2016 November 10--20, using the Los Cumbres Observatory (LCO) 1.0-m telescopes in Australia, South Africa and Chile.  These multi-site observations were requested for complete phase coverage of the orbit and the pulsations to allow a precise Fourier analysis.  However, only 18\% of the data was useful due to bad weather and scheduling restrictions.  For the usable LCO images, groups of four 5-s exposures obtained with the V filter were stacked prior to analysis and otherwise reduced as described above.

\subsection{Spectroscopy}
\label{sec:Spec}
Spectra were obtained using the ANU 2.3-m telescope at Siding Spring Observatory and the Wide-field spectrograph (WiFeS) \citep{dopita_wide_2007, dopita_wide_2010}.  Observations were taken over 8 nights in 2014 mid-November, with additional data obtained in 2016 mid-October. Observations covered most phases of the orbital cycle.  Spectra were obtained using the RT560 beam splitter and the B7000 and R7000 gratings for radial velocity (RV) determination.  Spectra for spectral classification were taken using the RT560/B3000 and RT480/U7000 combinations. The spectra were reduced with pyWiFeS, the data reduction pipeline specific for the WiFeS spectrograph \citep{childress_pywifes:_2014}.  Blocks of 4 consecutive TT Hor spectra were taken and alternated with Ne-Ar arc spectra and subsequently co-added in the pipeline. This reduced the effect of pulsations on the RVs, by integrating over more than half a pulsation period. 

Spectra were dominated by the primary star at all orbital phases, with the main absorption features being the Balmer series.  The orbit of TT\,Hor is inclined to $75^{\circ}$  (see Sect.\,\ref{sec:binary_model}) thus it is not possible to isolate the spectral features of the secondary component during the primary eclipse minimum.  There was no evidence of H$\alpha$ emission in these spectra.  

The spectral type was determined as A4IV by comparison of the TT\,Hor spectrum with spectral standards from \citet{gray_stellar_2009}. The standards of luminosity class IV gave a marginally better fit than luminosity class V. To determine atmospheric parameters, Local Thermodynamic Equilibrium (LTE) model atmospheres were taken from \textsc{atlas\,{\small 9}} (Castelli \& Kurucz, 2004) and synthetic spectra were calculated from these using \textsc{spectrum} (Gray \& Corbally, 1994). The blue data were continuum normalised and rectified prior to comparison to the synthetic spectra.The best fit to the data was with a synthetic spectrum with $T_{\rm eff}$ = 8800\,$\pm$\,200\,K, $\log g$ = 3.9 $\pm$\,0.3, [M/H]\,=\,0, microturbulance = $2$\,km\,s$^{-1}$ and $v \sin i$ = $50$\,km\,s$^{-1}$. 

Given the dominance of the Balmer features, RVs were extracted from the B7000 data rather than the R7000 data.  We used a custom Python interpretation of the two-dimensional correlation function, \textsc{todcor} (Zucker and Mazeh, 1994) to extract the RVs of both components.  The code uses template spectra that match the spectra of the two components of the binary.  Templates were generated from the synthetic spectra from \textsc{atlas\,{\small 9}} as described above and a grid-based search was used to determine $\alpha$, the contrast ratio of the two template spectra.  Maximum correlation was found using $\alpha$ =0.1, the parameters of $T_{\rm eff}$, $\log g$, $v \sin i$ = [8500--9000\,K, 4.0, $50$\,km\,s$^{-1}$] for the primary star and [4750--5000\,K, 3.5, $50$\,km\,s$^{-1}$] for the secondary star, both with solar metallicity.

The primary star parameters are consistent with the spectral typing given above, with RV measurements for this star being more reliable than for the secondary.  Uncertainties are lowest at quadrature phases (0.25 and 0.75), where the primary and secondary stars are at maximum velocity along our line-of-sight and spectral lines are Doppler shifted the most.  The maximum velocities set the limit on the masses of the two stars.  RV uncertainties at these phases for the primary and secondary components are no more than 1\,km\,s$^{-1}$ and 4km\,s$^{-1}$, respectively.  Standards stars analysed via \textsc{todcor} gave RV uncertainties of 2.5\,km\,s$^{-1}$.  We therefore add these uncertainties in quadrature to give uncertainties for the radial velocity semi-amplitudes of 2.7\,km\,s$^{-1}$ for \textit{K}$_1$ and 4.7\,km\,s$^{-1}$ for \textit{K}$_2$.
A table of RV measurements are given in the Appendix (Table~\ref{tab:refRV}).

\section{Binary Model}
\label{sec:binary_model}

\begin{figure}
    \includegraphics[width=\columnwidth]{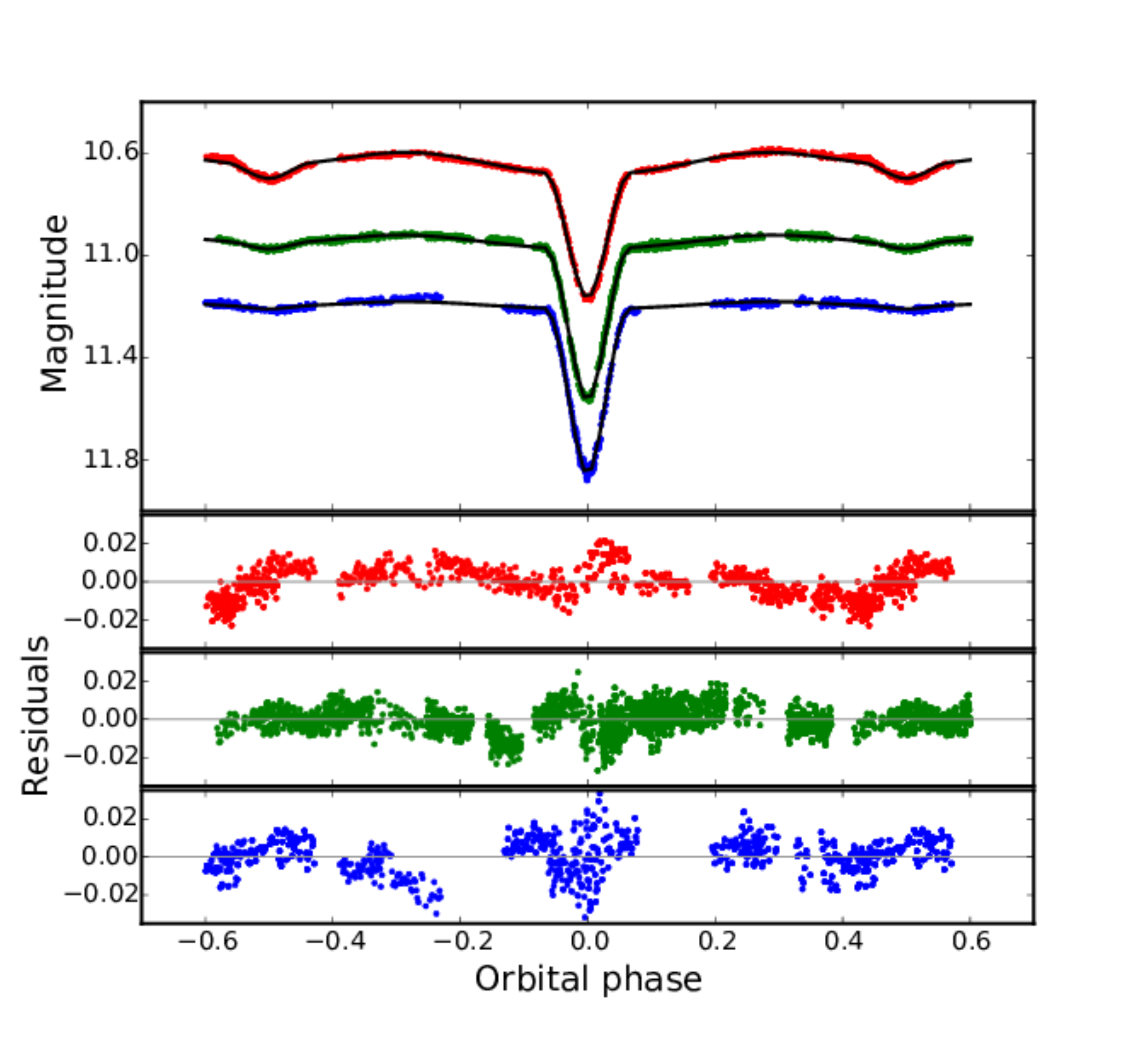} %Removed extension.
     \caption{Magnitude changes of TT\,Hor with orbital phase using different filters.  From top to bottom: Red = I filter, Green = V, Blue = B.  The final \textsc{phoebe} model for each filter are indicated by the black solid lines with the residuals for each filter plotted below.   The standard deviation for the magnitude of the check star was typically 0.005.}

    \label{BVI_Mags.eps}
 \end{figure}

Composite light curves for TT\,Hor in V, B and I from the 2014 to 2016 data are shown in Fig.~\ref{BVI_Mags.eps}.The system is brightest in the I passband and dimmest in blue with B$-$V $\approx 0.26$ during uneclipsed phases.  The light curves are typical of a semi-detached binary system.  The ingress and egress to and from the primary eclipse are deeper compared to those of the secondary eclipse, an effect which is most evident in the I data.  The secondary eclipse is also deeper in the I passband. These effects are indicative of a red secondary star having filled its Roche lobe.

The binary light curves in the three passbands and the radial velocity data were used together to determine the system model using \textsc{phoebe} \small{0.31}\textsc{a} \citep{prsa_computational_2005}.  \textsc{phoebe} is a modelling package for eclipsing binary systems based on the Wilson and Devinney code \citep{wilson_realization_1971}.  Our standardised data in three passbands maximise the power of the \textsc{phoebe} code to preserve the colour indices and this constrains the model to readily differentiate the effective temperatures of both components.

\begin{figure}

   	\includegraphics[width=\columnwidth]{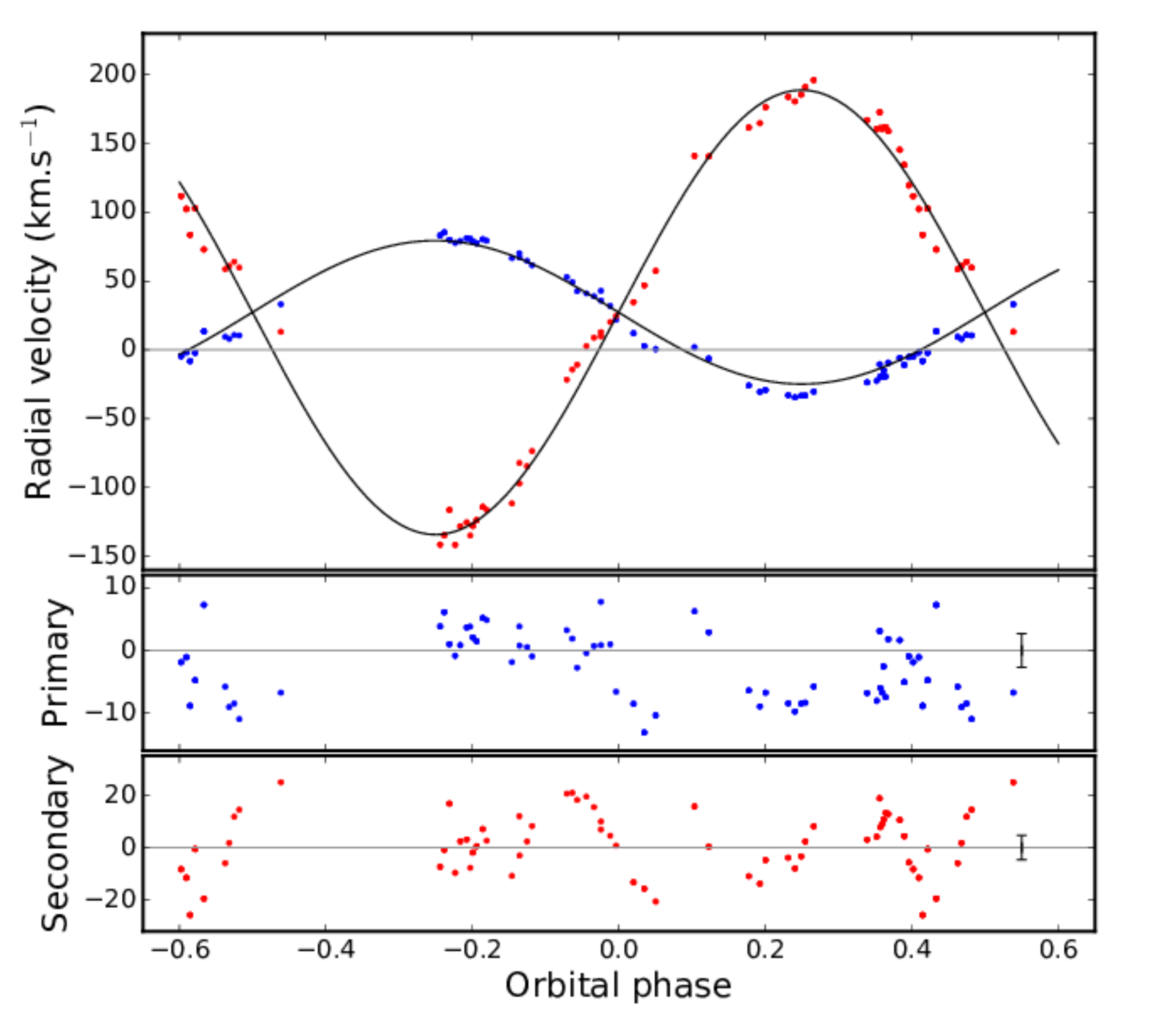}
     \caption{RV variations of the primary (blue) and secondary (red) stars with the orbital phase of the binary system.   The residuals remaining after the subtraction of the {\sc phoebe} model are shown below.  The error bars indicate the uncertainties at quadrature phases. }

    \label{RVS.eps}
\end{figure}

The final model of the system uses the radial velocities determined from the parameters detailed in Sect.\,\ref{sec:Spec}.  In addition, RV data gathered in 2016 November compared to those collected in 2014 October showed a consistent system velocity difference of 12.5\,km\,s$^{-1}$ (Table~\ref{tab:PHOEBE_model}) which could not be attributed to RV changes due to pulsations.  The 2016 data were corrected by this amount and combined with the 2014 data to give the final data set used (Fig.~\ref{RVS.eps}).

The difference in RVs between 2014 and 2016 is postulated to result from a third, low-mass object in a wide orbit about the binary system.  This is consistent with the predictions of \citet{tokovinin_tertiary_2006} that 96\% of binaries with orbital periods less than 3 days have tertiary companions.  The overall evolution of the system is affected by Kozai cycles with tidal friction as illustrated by \citet{eggleton_orbital_2001} for the semi-detached binary, $\beta$\,Per, the prototype semi-detached Algol system.   However, at the time of our observations, the Kozai cycles are no longer influencing the binary, as the perturbation caused by the distortion of the Roche lobe-filling component is much larger than the perturbation due to any third body. Of course, the third body is still having a gravitational effect on the binary as seen by the change in RVs with time.

\begin{table*}
	\caption{The parameters of TT\,Hor as determined by {\sc phoebe} from combined photometry and spectroscopy. Parameters that have purely statistical uncertainties from PHOEBE are marked with `$^{\wedge}$'.}
	\label{tab:PHOEBE_model}
	\begin{tabular}{lccr} % four columns, alignment for each
		\hline
		Parameter & Primary  & Secondary & System\\
        &(accretor)&(donor)&\\
		\hline
		Orbital period (days) & - & - & 2.608223 $\pm$ 0.000004\\
		Semi-major axis  ($R_{\sun}$) & - & - & 11.40 $\pm$ 0.07\\                                                         
		\textit{M}$_2$/\textit{M}$_1$ & - & - & 0.323 $\pm$ 0.019\\
        System velocity (km\,s$^{-1}$) & - & - & 27.2 $\pm$ 2.5\\
        2014 + 2016 & - & - &  \\
        2014 & - & - & 27.0 $\pm$ 2.5\\
        2016 & - & - & 15.3 $\pm$ 2.6\\
        Semi-amplitudes, \textit{K} (km\,s$^{-1}$) & 52.2 $\pm$ 2.7 & 161.7 $\pm$ 4.7 &  \\
        Inclination, (\textdegree) & - & - & 75.00 $\pm$ 0.01$^{\wedge}$\\
        \textit{M} ($M_{\sun}$) &	2.22 $\pm$ 0.17 &	0.72 $\pm$ 0.06  &\\
        \textit{R} ($R_{\sun}$)	& 2.05 $\pm$ 0.06	 &  3.26 $\pm$ 0.09 &\\
        \textit{T}$_\textit{eff}$ (K)	& 8800 $\pm$ 200	& 4627 $\pm$ 50 &\\
        log \textit{g}	& 4.16 $\pm$ 0.05	& 3.27 $\pm$ 0.05&  \\
        Surface albedo	& 1.0 (fixed)	& 0.761 $\pm$ 0.009$^{\wedge}$ &\\
        Gravity darkening coefficient &	1.0 (fixed) &	0.529  $\pm$ 0.009$^{\wedge}$ &\\
        Normalised surface potential, $\Omega$ &	5.912 $\pm$ 0.002$^{\wedge}$	& 2.495$^a$ &\\
        Log(L/$L_{\sun}$) &	1.35$\pm$0.05 &  0.64$\pm$0.03 &\\
        
		\hline
	\end{tabular}\\
$^a$Computed from masses and radii based on the assumption of Roche lobe overflow.
\end{table*}

The {\sc phoebe} minimisation fitting method of differential corrections was used throughout the modelling process. The shapes of the light curves are indicative of a semi-detached system and {\sc phoebe} was used in this mode. The orbital period ($P$) for the system was determined previously as 2.608204\,d \citep{moriarty_discovery_2013} and, with the additional 2014-2016 data, we refined this to $P$ = 2.608223 $\pm$  0.000004\,d.  Using this $P$ and the RV data, we derived and then fixed the semi-major axis, the mass ratio of secondary to primary ($q$\,=\,$m_2$/$m_1$) and the system velocity.  The maximum velocities at the critical quadrature phases give $q$. With uncertainties for the RVs conservatively set at no more than 5\,km\,s$^{-1}$ (3\% uncertainty), this gives an $\approx$10\% uncertainty in the final mass determinations. 

Importantly, the model of RVs during primary eclipse shows no clear evidence of the Rossiter-McLaughlin (RM) effect \citep{rossiter_detection_1924}.  During a primary eclipse, parts of the rotating surface of the primary star are hidden. This results in a redshift from the receding stellar surface and then a blueshift from the advancing surface compared to the orbital motion of the star. The spectral changes reveal the projected stellar rotation speed ($v \sin i$) and the angle between the stellar and orbital spins.  For the primary star RVs, an upper limit of 10 km\,s$^{-1}$ for the RM effect is seen which is consistent with the orbital and spin axes being aligned. 

The surface morphology of the stars depends on their Kopal surface potentials ($\Omega$) \citep{kopal_classification_1955} which were calculated by {\sc phoebe} from the radii and mass ratios, assuming a circular orbit and synchronous rotation.  The potentials were fixed initially from the radii of the two stars as determined from the photometry.  When the primary $\Omega$ was set as a free parameter to refine the final model, the resultant normalised value changed by only 0.002.  The greater surface potential of the primary star compared to the secondary reflects the smaller radius of the primary star which is bound within a separate equipotential surface from the lobe-filling secondary. 

The best fit model was searched for with the primary $T_{\rm eff}$ set as 8800\,K and the secondary $T_{\rm eff}$, together with the inclination angle $i$, set as free parameters. Models were also computed with $T_{\rm eff}$ set as 9000\,K or 8600\,K.  This change only affected the temperature of the secondary star, and enabled the temperature uncertainty to be propagated. Limb darkening coefficients for each filter and temperature were determined using the linear cosine law and \citet{van_hamme_new_1993} tables and then fixed.   For the primary star, the albedo and gravity darkening power-law coefficients for mean intensity are set to 1.0, as appropriate for radiative envelopes. This disregards the likely presence of a very thin convective envelope in the primary component.  For the secondary star, they were set as free parameters and minimised at 0.529 $\pm$ 0.009 and 0.761 $\pm$ 0.009, respectively, indicating a convective envelope as expected for this cooler component.

\begin{figure}

   	\includegraphics[width=\columnwidth]{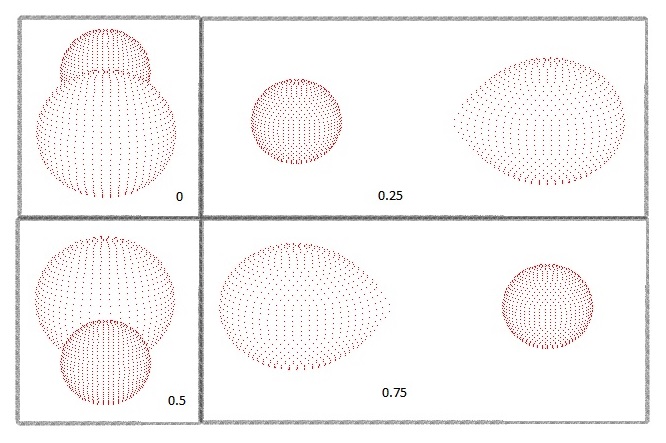}
     \caption{\textsc{phoebe} representation of the shape models during the phase changes of TT\,Hor.  From left to right, top to bottom: phase\,-0 (primary eclipse); phase-   0.25; phase- 0.5 (secondary eclipse); phase- 0.75 (-0.25).}

    \label{AllPhases.jpg}
 \end{figure}

The final {\sc phoebe} model as fitted to the data for each filter band, together with the residuals, are shown in Fig.~\ref{BVI_Mags.eps}.  The residual magnitudes for each filter band are between 0.2 and 0.3\% with the B filter having the highest residuals.  The geometry of the TT\,Hor system at different phases is shown in Fig.~\ref{AllPhases.jpg}. The primary star has the smaller diameter with the secondary star swollen to fill its Roche lobe.  The $75^{\circ}$ inclination of the orbital axis is evident at the eclipse phases when both stars are visible.  The models shows no indication of eccentricity and is consistent with a circular orbit.

Pulsations are clearly evident throughout the orbital cycle, particularly in the B and V filters, and contribute to the residuals from the {\sc phoebe} model.  The pulsations and other residuals were subtracted from the V data model and the main frequencies of the pulsations were then determined using \textsc{period\,{\small 04}} \citep{lenz_period04_2005}.  \textsc{period\,{\small 04}} is a discrete Fourier transform algorithm designed specifically for the statistical analysis of large time series containing gaps. However, the gaps in our data are mostly too large and the data sets from each night were analysed separately (Section\,\ref{sec:pulsation}).  Some data sets are also too short for Fourier analysis and were not included in the final pre-whitened orbital data.

\begin{figure}

    \includegraphics[width=\columnwidth]{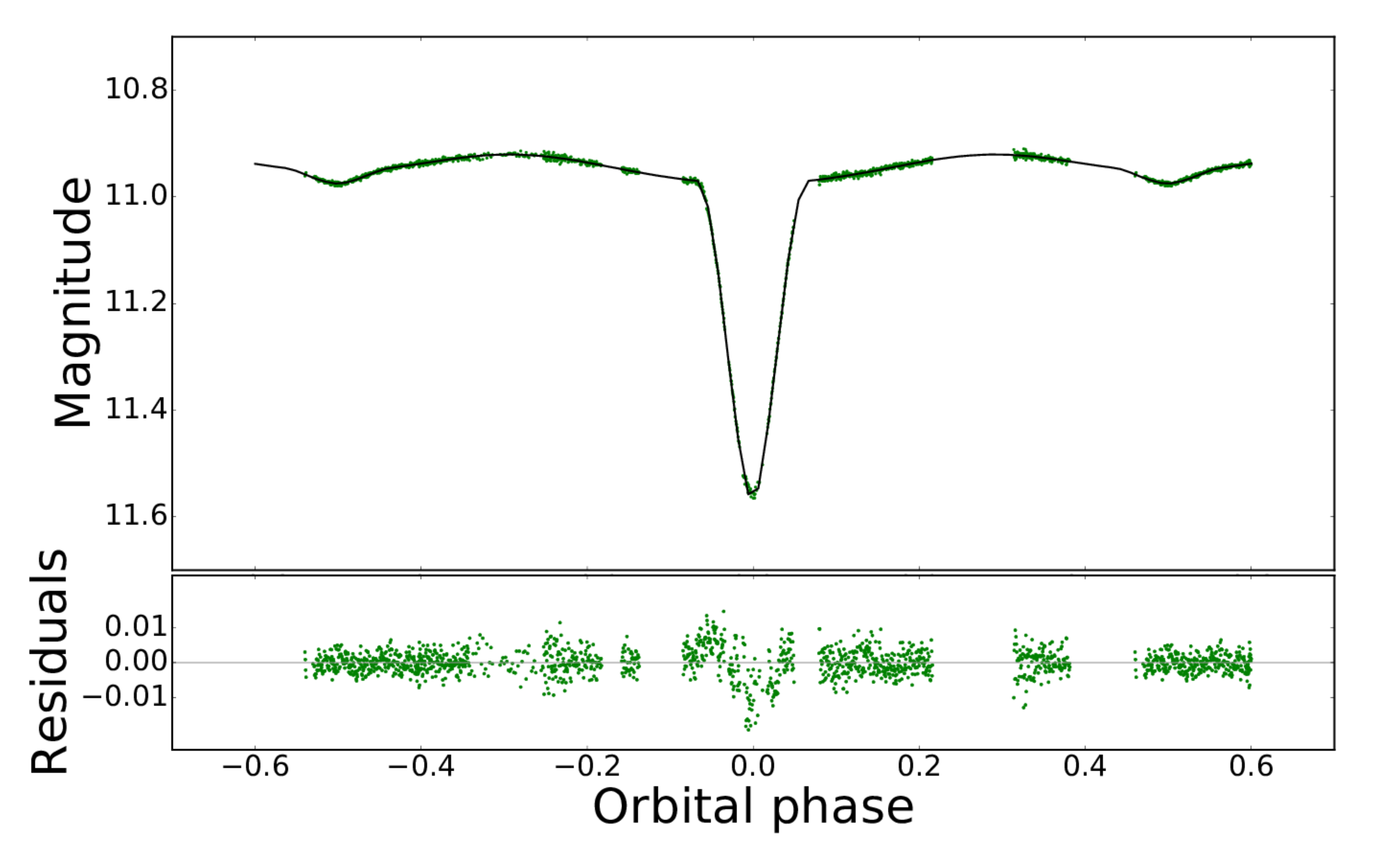}
     \caption{Magnitude of TT\,Hor (green dots) in V filter after pre-whitening compared to the {\sc phoebe} model (black line). The residuals from the model are shown below.}

    \label{pre-whitenedVA.eps}
 \end{figure}

The model for the frequency fit was then subtracted from the original data sets to produce a new phase-folded light curve without pulsations (Fig.~\ref{pre-whitenedVA.eps}).
 
The residuals for this prewhitened model (minus pulsations) are reduced significantly to less than 10\,mmag for all phases except for parts of the primary eclipse.  The higher residuals in the latter are explained by the difficulty of determining the exact times of eclipse minima.  Only the primary star pulsates but because of the inclined orbital axis, the eclipses are partial and pulsations are visible at both eclipse minima.

  The parameters for the final binary model are given in Table~\ref{tab:PHOEBE_model}. The radius uncertainty is dominated by the velocity uncertainty, which is integrated to give the physical orbital scale. For parameters that have purely statistical uncertainties from {\sc phoebe}, we have clearly marked them as such with `$^{\wedge}$'.  These uncertainties come directly from the PHOEBE model fit, which assumes uncorrelated observational errors. Given that our photometric uncertainties are correlated on scales at least equal to the residual pulsation frequencies at $\approx$40 d$^{-1}$  or 6 observations, these uncertainties are scaled by a factor of at least two.

The {\sc phoebe} model confirms the semi-detached nature of TT\,Hor, consisting of a 2.22 $\pm$ 0.17-M$_{\sun}$ primary (hereafter termed the accretor) and a 0.72 $\pm$ 0.06-M$_{\sun}$ secondary (hereafter termed the donor).  From the temperatures and luminosities, the donor is likely a H-shell burning, K-giant. From the luminosity of the accretor, we estimate the distance of TT\,Hor to be $\approx$ 800\,pc. This estimate uses the V magnitude (10.98) at secondary eclipse and a bolometric correction of zero for an A-type star \citep{flower_transformations_1996}. Because of the inclination of the orbital axis, the K-giant's contribution to the V magnitude is not negligible at this phase, so the distance is probably under-estimated.  Also no correction was made for extinction which, though slight for our results, would work to over-estimate the distance.  \red{Recently released Gaia DR2 data give a distance for TT Hor of 786 $\pm$ 18 pc \citep{gaiacollaboration2018}, thus our estimated distance is within Gaia DR2 uncertainties. }

The donor has evolved to fill its Roche lobe (radius = 3.26\,$R_{\sun}$ and is transferring mass to the accretor.  The model is consistent with TT\,Hor being an Algol-type binary.  In these binaries, the larger mass star, now the donor, evolves the fastest to its sub-giant phase and fill its Roche lobe.  Mass is transferred to the (originally) less massive component, the accretor.  We observe the system when the accretor is now the more massive star having gained considerable mass from the donor.

\section{Pulsation analysis}
\label{sec:pulsation}

Pulsations seen in data taken with the B and V filters are of similar amplitudes but are barely visible in the I data.  Our V data are the most abundant and were used for pulsation analysis. The best data sets are those taken on nights of excellent seeing and >4\,h duration. The orbital phase changes for each night's data need to be removed prior to pulsation analysis.  Initially, an attempt was made to remove the relevant phase by subtraction of the {\sc phoebe} model but matching the phase exactly to data sets other than those used in the phase folded light curves was difficult.  The subtraction of the binary model left an obvious inflection point with some data above and the rest below the zero axis.  Finally a low order polynomial was fitted to each 'good' data set with removal of any points at the beginning and end of each data set if the polynomial fit was inconsistent.

The best data sets from 2013 (10 nights, 64.5\,h) were combined for Fourier analysis using {\sc period\,{\small 04}}, as were those from 2014 (5 nights, 35.5\,h).  Observations obtained with the LCO in 2016 were too intermittent and not in themselves of sufficient cadence for Fourier analysis.  However, two data sets of 8 and 7 hr duration were obtained on 2016 November 12 and 17, respectively, from Murrumbateman using the 350-mm Meade Schmidt-Cassegrain telescope and SBIG STT 1603 camera using the V filter only.  When combined with the LCO data (30\,h) between these two dates, Fourier analysis was successful.  The frequencies extracted for the three years are given in Table~\ref{tab:frequencies}.   The four frequencies have different amplitudes for different years as indicated by their order of extraction from the Fourier transform. 
Extracting more than four frequencies resulted in either an extraction of sidelobes of the dominant four, or frequencies extracted in the low frequency, systematic noise floor of the power spectrum.

A combined analysis of all three years is also given in Table~\ref{tab:frequencies}. This latter Fourier analysis suffers from multiple aliases but gives more reliable amplitudes.  However, the highest and second-highest peak amongst the 1\,yr$^{-1}$ aliases are not distinguishable at our signal-to-noise values. For this reason, we conservatively assign 1\,yr$^{-1}$ (or 0.003\,d$^{-1}$) as the uncertainty in the combined frequency analysis. 
 \begin{table*}
   	\centering
  	\caption{Frequency analysis of data sets from 2013, 2014, 2016.  These data were either analysed individually for each year or combined into one data set covering 2013 to 2016.  The errors for the individual data sets are those given by {\sc period04} for Monte Carlo simulations. }

 	\label{tab:frequencies}
	\begin{tabular}{lcccccccc} % ninecolumns, alignment for each
		\hline
		 & \multicolumn{2}{c}{Combined 2013 - 2016}& \multicolumn{2}{c}{2013} & \multicolumn{2}{c}{2014} & \multicolumn{2}{c}{2016}\\
        & Frequency  &Amplitude & Frequency  &Amplitude  &Frequency  &Amplitude  &Frequency  &Amplitude \\
        &(d$^{-1}$)& (mmag)& (d$^{-1}$)& (mmag) & (d$^{-1}$)& (mmag) &(d$^{-1}$)& (mmag)\\
		\hline
		\textit{f}$_1$& 39.907 $\pm$ 0.003& 3.8 $\pm$ 0.3& 39.902 $\pm$ 0.001& 5.1$\pm$ 0.4 & 38.073 $\pm$ 0.001 & 3.4 $\pm$ 0.2 & 39.920 $\pm$ 0.002 & 4.8 $\pm$ 0.2\\
		\textit{f}$_2$& 38.075 $\pm$ 0.003 & 2.8 $\pm$ 0.5  &38.074 $\pm$ 0.001 & 4.1$\pm$ 0.3 &34.308 $\pm$ 0.001 & 3.1 $\pm$ 0.2 &42.283 $\pm$ 0.003 & 2.4 $\pm$ 0.3\\
		\textit{f}$_3$&34.310 $\pm$ 0.003 & 2.2  $\pm$ 0.6 & 34.293 $\pm$ 0.001 & 2.2$\pm$ 0.3 & 39.903 $\pm$ 0.001 & 2.6 $\pm$ 0.2 & 38.067 $\pm$ 0.003 & 2.6 $\pm$ 0.3\\
      \textit{f}$_4$ &42.301 $\pm$ 0.003 & 2.0 $\pm$ 0.1& 42.307 $\pm$ 0.001 & 2.3$\pm$ 0.3 & 41.198 $\pm$ 0.001 & 1.9 $\pm$ 0.2 &34.325 $\pm$ 0.006 & 2.0 $\pm$ 0.3\\
		\hline
	\end{tabular}
\end{table*}

Our ground-based data do not allow the detailed pulsation analysis possible with {\it Kepler} and CoRoT data but as all four frequencies are consistently present over the three years of observations, they are considered authentic.   Although not ideal, they are useful for a preliminary identification of the modes (see Sect.\,\ref{ssec:mode-id}).   

As TT\,Hor is inclined at 75$^{\circ}$ to our line of sight, the eclipses are partial and a spatial filtering effect is expected during primary eclipse \citep{moriarty_discovery_2013}.  This effect can lead to an amplitude increase in some oscillation modes, and could lead to modes being observed that are not observable outside of eclipse. For the V data (of highest signal to noise), we attempted to search for such an effect. Only one reliable frequency could be found in common amongst our 7 primary eclipses, of 37.7\,d$^{-1}$, with a dispersion of 0.8\,d$^{-1}$, and an amplitude of 9.8 mmag with a dipsersion of 2.8 mmag. Given that a primary eclipse only lasts approximately 11 pulsation cycles, the identity of the mode or combination of modes which produced this apparently higher amplitude is unclear.

\section{Modelling}
\subsection{Evolutionary models}
An evolutionary model to match the \textsc{phoebe} binary model is essential to fully understand the TT\,Hor system.  We used the \textsc{mesa} stellar evolutionary code, version r-8845 \citep{paxton_modules_2011,paxton_modules_2013, paxton_modules_2015} to generate evolutionary models from different initial conditions of mass and orbital period.  A  grid search was performed with the total initial mass for the binary system set at 2.931 $M_{\sun}$ (total mass of the TT Hor system to four significant figures), the initial mass of the donor varied from 1.6 to 1.9 $M_{\sun}$ and the initial orbital period varied from 2.6 to 2.9 days.  A final model was subsequently found with the donor mass varied from 1.61 to 1.63 $M_{\sun}$ and that of the accretor from 1.301 to 1.331 $M_{\sun}$. For computation of the evolutionary models, an additional decimal place for the masses of the binary stars was used for better discrimination between these models for all parameters.  

The default solar values for hydrogen (X), helium (Y) and the metal fraction (Z = 0.02) were used with convective-core overshoot in both the donor and accretor.  As the initial masses of the donor and accretor for these final models were less than 2 $M_{\sun}$, $\alpha$$_{ov}$ was set at 0.1 as determined from  \citet{claret_dependence_2017}, for a 1.6-M$_{\sun}$ star.  The other core-overshoot parameters for \textsc{mesa} were set as f0 = 0.008 (a non-zero value for the edge of the convective zone in terms of scale height)  and f = 0.01 (the pressure scale height of overshoot above the convective zone).   

Mass transfer can be either conservative or non-conservative. With non-conservative mass transfer, mass is lost from the system and either an accretion disc or a `hot-spot' associated with the accretor may drive the outflow of mass \citep{deschamps_non-conservative_2015}.  Asymmetries in the binary light curves of RZ Cas \citep{rodriguez__2004} and KIC 5621294 \citep{lee_kepler_2015} have been attributed to the existence of hot-spots.  Our \textsc{phoebe} model of the TT\,Hor system did not require any hot (or cool) spots to explain the shapes of the light curves and no H$\alpha$ emission was seen in the spectra.  No other evidence suggested that mass transfer is anything but conservative.  Indeed, low-mass binary systems (total mass $\leq$ 3 $M_{\sun}$) have been successfully modelled by {  \citet{kolb_comparative_1990} with mass transfer occurring via a fully conservative mechanism (i.e. without systemic mass loss).  In these systems the rate of mass transfer is in the order of 10$^{-10}$ $M_{\sun}$/yr.  Given the semi-detached nature of the system and the masses of the component stars, mass transfer was assumed to be conservative and we modelled this with the ``Kolb'' binary control in {\sc mesa}, following the Kolb and Ritter formalism. 

The properties of the \textsc{mesa} models closest to the observed binary system are shown in Table \ref{tab:MESA_models}.  Using solar metallicity (Z = 0.02), initial donor mass = 1.61 $M_{\sun}$; accretor mass = 1.321 $M_{\sun}$; initial period = 2.77\,d, we are able to construct a \textsc{mesa} model (Model A) that matches our \textsc{phoebe} model for the masses of both components.  While all other parameters differed by about 1\% at this mass, the temperature and radius of the accretor differed by +8\% and -4\%, respectively.  These differences are greater than the uncertainties given by the \textsc{phoebe} model ($\approx 2\%$ and 3\%, respectively) and too great to be disregarded and suggested that the metallicity of the system might be greater than solar.   Therefore, Z was varied from 0.02 to 0.036 to search for a {\sc mesa} model that gives a lower temperature and higher radius for the accretor.  The closest fitting model (Model B) was for Z = 0.033; initial donor and accretor mass as above; initial period = 3.05\,d.  

Because of the uncertainties in our RV measurements, there is a possibility that the masses of the two components of TT\,Hor are up to 10\% smaller than calculated by \textsc{phoebe} modelling.  Therefore an additional \textsc{mesa} model (Model C) with the same mass ratio (0.323) as the \textsc{phoebe} model but with the final mass for the accretor and donor reduced by 10\% (to 2.016 and 0.65 $M_{\sun}$, respectively) was also determined using solar metallicity.   With the initial donor mass\,= 1.44\,$M_{\sun}$; accretor mass = 1.226 $M_{\sun}$; initial period\,=\,2.95\,days, the temperature of the accretor and donor both matched that of the \textsc{phoebe} model.   Radii, luminosity and log\textit{g} were within the 10\% error margin.

 \begin{table*} 
	\centering
	\caption{Comparison of \textsc{phoebe} binary model for TT\,Hor and \textsc{mesa} evolutionary models with different initial conditions of mass, orbital period and Z. Evolutionary models for single stars of corresponding masses and Z to the accretor are also given. For single star, Model A, the parameters are given for this model which match both \textsc{mesa} Model A and the \textsc{phoebe} binary model. }   
	\label{tab:MESA_models}
	\begin{tabular}{lccccccccc} % ten columns, alignment for each
		\hline
 		Parameters& \textsc{phoebe} & \multicolumn{4}{c}{\textsc{mesa}} & \multicolumn{4}{c}{Single Star}\\
        &Model &Model A &  Model B & Model C &\red{ Model D}  &Model A &  Model B & Model C  & \red{Model D}\\
       
		\hline
        Initial mass&	- & 1.61 &	1.61 &	1.44 & 1.548&&\\
        ~(donor) ($M_{\sun}$ )&&\\
        Initial mass &	- &	1.321 &	1.321 &	1.226 &1.287 &&\\
        ~(accretor) ($M_{\sun}$)&&\\
        Initial orbital \textit{P} &	-	& 2.84 & 3.05 &	2.95 & 3.00 &&\\
        ~(days)&&\\
        Z fraction &	-	& 0.02	& 0.033 &	0.02& 0.0281& 	0.02	& 0.033	& 0.02&0.0281\\
        Final orbital \textit{P}  & 2.608223 & 2.60353 & 2.61282&	2.60349 & 2.56812 &-&- &- &-\\
        ~(days) &&\\
		Semi-major axis  & 11.40 $\pm$ 0.07	&11.388	&11.422&11.040	& 11.166&-& - & - & - \\                                                           ~(R$_{\sun}$) &&\\
		
        Age (Gyr) & - &	2.03	& 2.58	& 2.96	&2.72&	0.50	& 0.30	& 0.38&0.29\\ 
       \textit{M}$_2$/\textit{M}$_1$ & 0.323	& 0.323	& 0.323 &	0.323 &0.265& - &- &-&-\\ 
        \multicolumn{2}{l}{\textbf{Primary (accretor)}}&&\\
        \textit{M} ($M_{\sun}$)  &	2.22 $\pm$ 0.17 &	2.217 &	2.216 &	2.018 &	2.153  & 	2.216 &	2.216 &	2.016 & 2.153 \\
        \textit{R} ($R_{\sun}$)	& 2.05 $\pm$ 0.06 &	1.968 &	2.076 &	1.951 & 2.014& 	1.869/2.435 &	2.176 &	1.862&2.014 \\
        \textit{T}$_\textit{eff}$ (K)	& 8800 $\pm$ 200 &	9571 &	8787 &	8807 &	8827	& 9585/8805	& 8797 &	8820& 8881\\
        log \textit{g}	& 4.16 $\pm$ 0.05	& 4.20 &	4.15 &	4.16 &	4.16 &	4.24/4.01 &	4.11 &	4.20& 4.16\\
         Log(L/$L_{\sun}$)	& 1.35 \red{$\pm$ 0.05} &	1.46 &	1.36 & 	1.31 & 1.34 &	1.42/1.51 &	1.41 &	1.28 & 1.36\\
        \multicolumn{2}{l}{\textbf{Secondary (donor)}}&&\\
        \textit{M} ($M_{\sun}$) & 0.72 $\pm$ 0.06 & 0.714 &	0.715 &	0.648 &	0.682&&\\
        \textit{R} ($R_{\sun}$) & 3.26 $\pm$ 0.09 &	3.254 &	3.257 &	3.150 &	3.183 &&\\
        \textit{T}$_\textit{eff}$ (K) & 4627 $\pm$ 50 &	4712 &	4571 &	4666 &	4593&&\\
        log \textit{g} & 3.27 $\pm$ 0.05 &	3.27 &	3.27 &	3.25 &	3.27&& \\
        Log(L/$L_{\sun}$) & 0.64 \red{$\pm$ 0.03 }&	0.67 &	0.62 &	0.63 &	0.61 && \\
        
		 \hline
	\end{tabular}
\end{table*}

\red{Models B and C both fit the \textsc{phoebe} binary model within uncertainties.  However, a scaling factor was needed for identification of the modes from the \textsc{mesa} models (refer to Subsection~\ref{ssec:mode-id}). Therefore, a further model (Model D) was computed which did not require any scaling factor for mode identification of the observed frequencies.  For Model D, a lower total mass (3.3\%) and a slightly higher Z (0.0281) are needed compared to Model A, ie a model with parameters between those for Models B and C.  }

Models for the evolution of single stars with the corresponding masses and Z fraction, as the accretor were also constructed with \textsc{mesa} as shown in Table~\ref{tab:MESA_models}.  The single stars are considerably younger, 0.3 to 0.5\,Gyr, compared to the binary system, 2 to 3\,Gyr.   

The evolutionary tracks of the two binary components together with those of single stars of the same mass and metallicity as the accretor are shown in an HR diagram in Fig.~\ref{HR1.eps}  The single stars are crossing the main sequence, whereas the accretor in each model is tracking up the main sequence as it gains mass.   The donor in each model has moved off the main sequence, having evolved to its subgiant phase, filled its Roche lobe and has lost $\approx44$\% of its original mass to its binary partner. \red{The black stars indicate where rapid mass transfer begins with the corresponding ages being:- Model A, 1.93\,Gyr; Model B, 2.50\,Gyr; Model C, 2.82\,Gyr; Model D, 2.63\,Gyr.  The open circles locate the current evolutionary states of each star on the HR diagram.}  For the single star with solar metallicity, \red{Model A,} and mass = 2.216 ($M_{\sun}$), two conditions in the $T_{\rm eff}$/$\log $($L$/L$_{\sun}$) parameter space are equivalent to either the binary accretor Model A \red{(left) or Models B and C (right)}.

\begin{figure*}

 	\includegraphics[width=\textwidth]{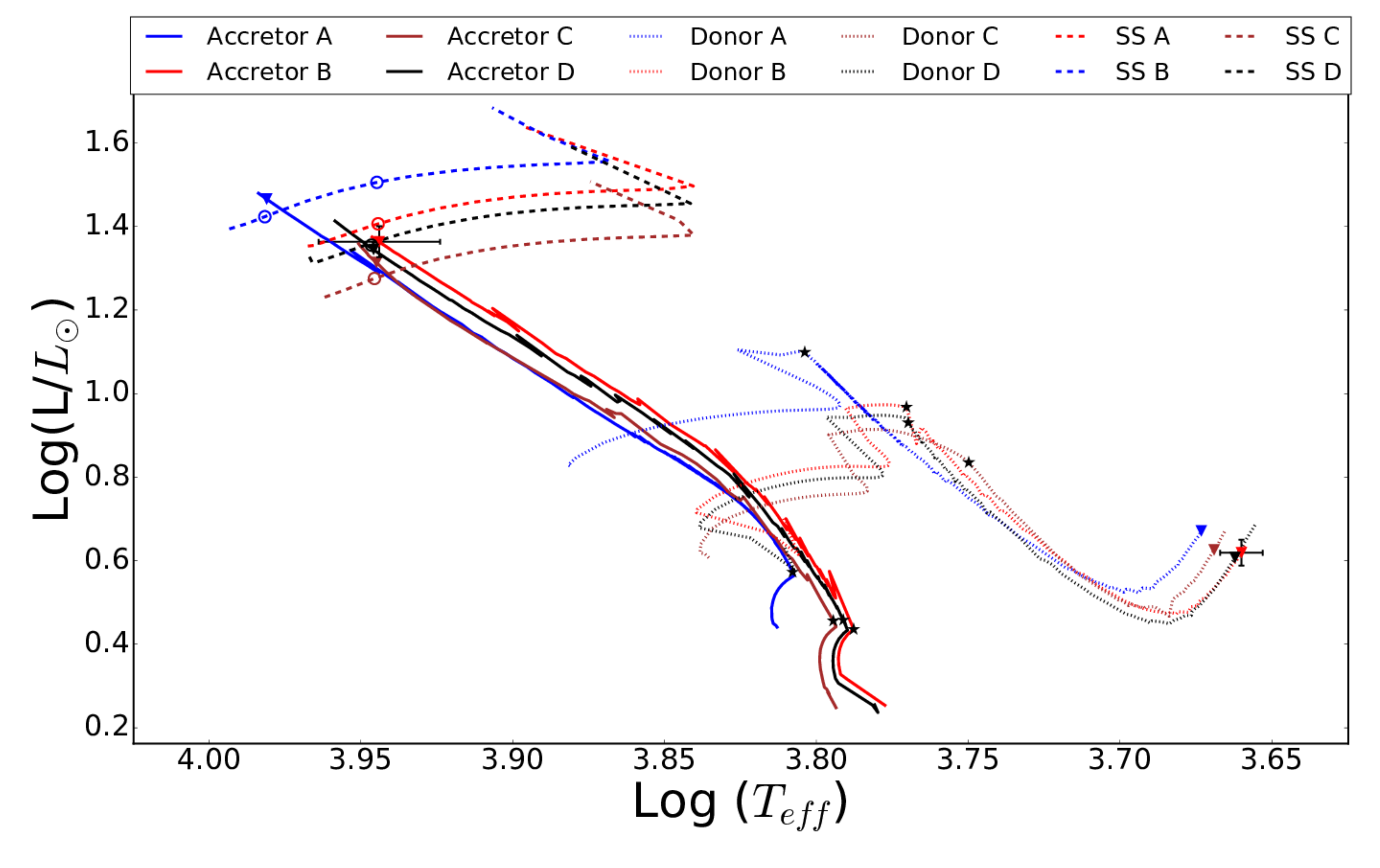}
    \caption{\red{HR diagram showing the evolutionary tracks of \textsc{mesa} models for the accretor and donor with different initial conditions as given in Table~\ref{tab:MESA_models}. Tracks for single stars (SS) of the same final mass as the accretor for each model are also shown. The inverted triangles indicate the current evolutionary state of each star. Error bars are shown for the accretor and donor from the \textsc{phoebe} binary model. The black stars indicate start of rapid mass transfer.    The open circles indicate the equivalent evolutionary conditions in the $T_{\rm eff}$/$\log $($L$/L$_{\sun}$) parameter space for the single stars. }} 

   \label{HR1.eps}
 \end{figure*}

Evolutionary Model A is rejected as the temperature of the accretor is too high (>8\%) when the accretor's evolutionary track matches the other parameters of the \textsc{phoebe} binary model as indicated by the inverted blue triangle. To reduce the temperature of the accretor required either an increase in metallicity (Models B \red{and D}) or a decrease in the final mass of the accretor (Models C \red{and D}).  While these \red{three} evolutionary models can be matched to the \textsc{phoebe} model within uncertainties, \red{Models B and D are} a closer match to the masses and radii of both accretor and donor as shown in Table ~\ref{tab:MESA_models}.  \red{For example,} the mass and radius of the accretor for Model B differ from the binary model by 0 and 1.5\%, respectively, whereas, for Model C, the differences are 8.9 and 4.6\%.  Similar differences are seen for the donor.   We, therefore, present more details for Model B with the metallicity for TT\,Hor being slightly greater than solar at [M/H] = 0.22.  Model C \red{and D have} similar properties for the parameter spaces given in Fig.~\ref{Age.eps} and Fig.~\ref{He4.eps}.

Considering Model B with [M/H] = 0.22, mass transfer from the donor (initial mass = 1.61\,$M_{\sun}$) to the accretor (initial mass = 1.321 $M_{\sun}$) is zero for the first 2.48\,Gyr and then rapidly increases to an average rate of 2.58\ $\times$10$^{-9}$ $M_{\sun}$yr$^{-1}$ as the donor evolves to fill its Roche lobe and the binary separation reduced to 7.9 $R_{\sun}$ (Fig.~\ref{Age.eps}).  Mass transfer is conservative throughout as the accretion rate is much lower than the Eddington limit because of the slow subgiant evolution.  When the two components are of equal mass, at 2.54\,Gyr, a rapid separation occurs to reach their final separation seen today (see \citet{eggleton_evolutionary_2006} and footnote\footnote{A description of the dependence of the binary separation on mass ratio and envelope properties can be found at \url{https://joe-antognini.github.io/astronomy/mass-transfer-modes}}).

\begin{figure}

 	\includegraphics[width=\columnwidth]{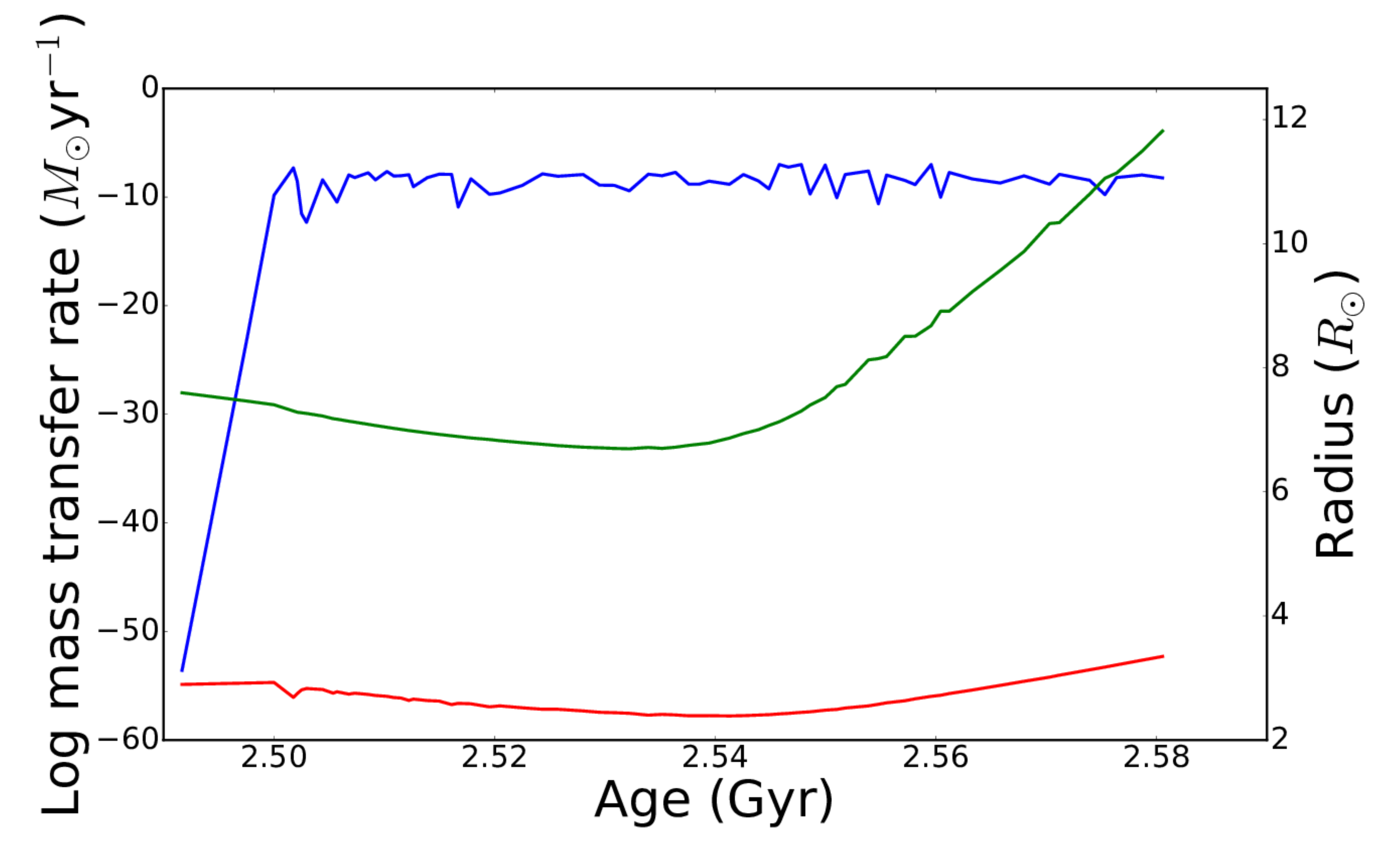}
    \caption{Change in  mass transfer with age (blue line) and the separation of the accretor and donor (green line) for Model B.  The radius of donor is shown in red.}

   \label{Age.eps}
 \end{figure}

The surface layers of the accretor are slightly enriched in $^4$He compared to a single star of the same mass and metallicity as shown in Fig.~\ref{He4.eps}.  This is expected from the accretion of nuclear-processed material gained from the donor during 80 Myr of mass transfer. 

\begin{figure}

 	\includegraphics[width=\columnwidth]{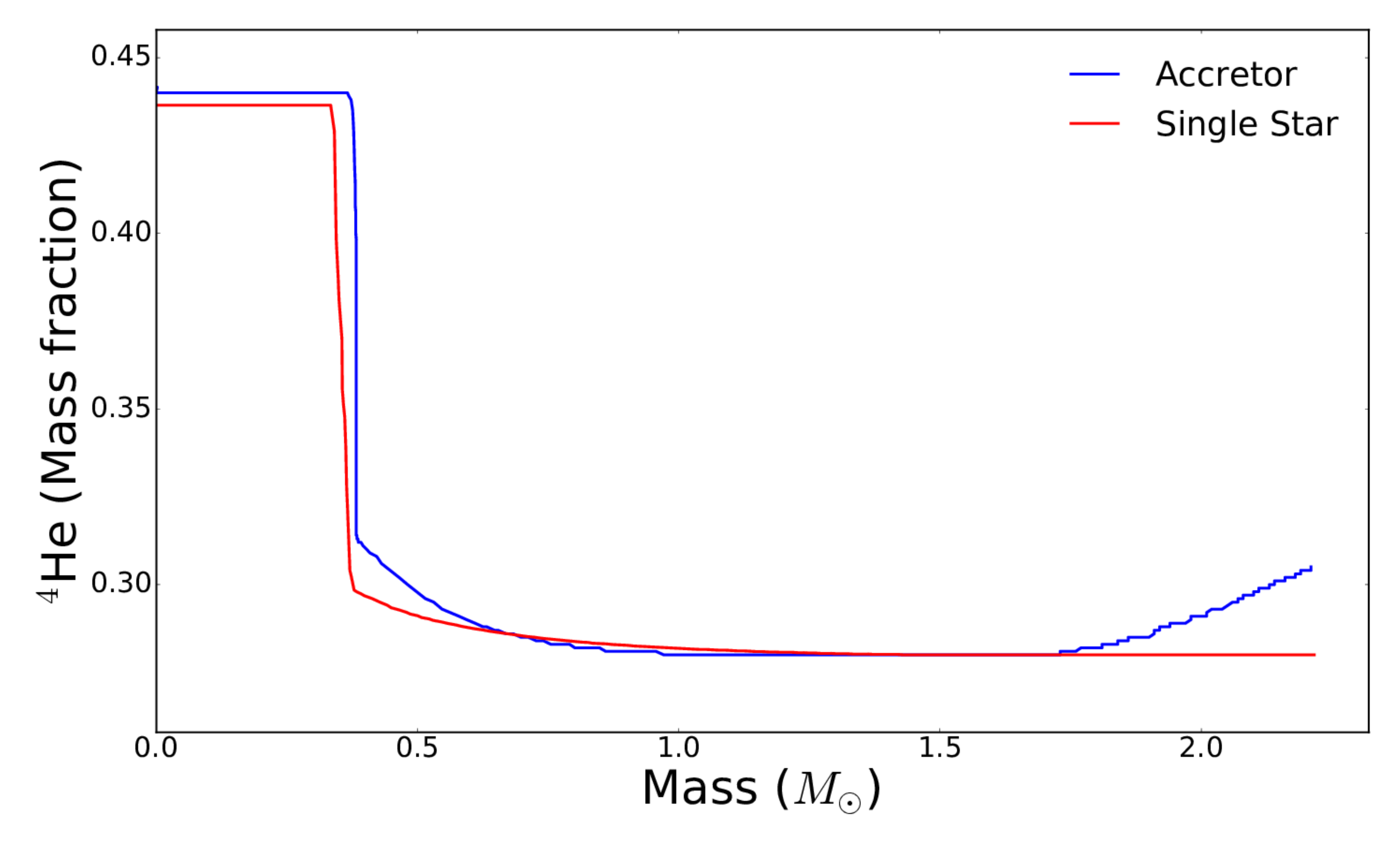}
    \caption{Change in $^4$He mass fraction across the mass profile of a 2.216 $M_{\sun}$ star with Z = 0.033.  The accretor is shown in blue and a single star of similar parameters in red. }

   \label{He4.eps}
 \end{figure}

\subsection{Mode identification}
\label{ssec:mode-id}

From the short-duration, ground-based observations, we were only able to identify four reliable frequencies \red{(refer to Table~\ref{tab:frequencies})} for which we would like to identify the corresponding oscillation modes. Our \textsc{phoebe} model of the TT\,Hor binary system  matches the \textsc{mesa} evolutionary Model B well (within 1.5\% in all parameters).  In addition, the binary model shows no indication of orbital eccentricity, and the RM effect is minimal and consistent with the orbital and spin axes being aligned.  We also know the rotation period of the pulsator, so we considered a preliminary investigation of the pulsation properties a worthwhile exercise.  When TESS data become available a more rigorous analysis and identification of the pulsation modes will be undertaken.

\red{We searched for the observed frequencies in models computed by \textsc{gyre} \citep{townsend_gyre:_2013}, the asteroseismology extension of \textsc{mesa}.}  \textsc{gyre} solves first-order equations describing perturbations to the boundary conditions at each interval of time using the Magnus Multiple Shooting numerical scheme. We used this code to calculate the simple linear adiabatic frequencies for the accretor and single star profiles from their \textsc{mesa} evolutionary models for Z = 0.033 (Model B).  We set the rotation rate in the \textsc{gyre} code using the orbital period of the binary at 2.788$\times$10$^{-5}$\,rad\,s$^{-1}$.  We assumed the accretor rotates uniformly, with no adjustment for its slightly oblate shape. We also computed the non-adiabatic and quasiadiabatic frequencies but the corrections are very small ($\leq$ 0.01\%) and we are unable to discriminate between the different perturbation effects using our current data.  Additionally, \textsc {gyre} neglects perturbations to the turbulent pressure, which is known to be an additional driving mechanism for pulsations in at least some $\delta$\,Sct stars \citep{antoci_role_2014,xiong_turbulent_2016} This mechanism, if operative for the TT~Hor accretor, would not be modelled by \textsc {gyre}.

\red{Frequency models for the accretor in Model B could be matched to the observed frequencies after applying a slight scaling factor of 1.027 (Table~\ref{tab:modeID} and Fig.~\ref{Echelle.eps}). } The scaling factor applied to the frequencies is 2.7\%, and is essentially a 5.4\% density scaling which is within the uncertainties in radius and mass of the accretor given by the \textsc{phoebe} binary model (Table~\ref{tab:PHOEBE_model}).   To find equivalent frequencies in the single star \red{matching Model B }, a scaling factor of 1.102 was applied. For such a substantially different model, the maximum difference to the accretor model frequency was only 0.09\,d$^{-1}$. \red{Mode identification for Model C accretor and single star gave essentially the same modes after applying a scaling factor of 0.981 and are not presented here.   }

This non-standard procedure was used because of the complexity of fine-tuning the input for the \textsc{mesa} binary evolutionary models.  
\red{Small changes in metallicity of tens of percent should change the radius through opacity changes, in turn changing the density and scaling the frequencies. To confirm that this approach is reasonable, and also to 
find a fully self-consistent model that fits our data, a third \textsc{mesa} model, Model D, was calculated so that the calculated frequencies from Gyre matched the observed frequencies without the need for a scaling factor.  These identified modes are also presented in Table~\ref{tab:modeID} and displayed as an \'echelle diagram in Fig.~\ref{Echelle.eps}. For a single star equivalent to the accretor in Model D, no scaling again achieves the same mode identification. }

Fig.~\ref{Echelle.eps} shows \'echelle diagrams identifying the observed modes using linear adiabatic calculations for the accretor \red{ of Models B and D }using a large separation of $\Delta$$\upsilon$\,=\,5.2\,d$^{-1}$.  We can compare our large separation value to that obtained for seven binary systems by \citet{garcia_hernandez_observational_2015}. Our analysis is well within the error bars for the large separation-mean density relation given in Figure 1 by these authors.

\begin{figure}

 	\includegraphics[width=\columnwidth]{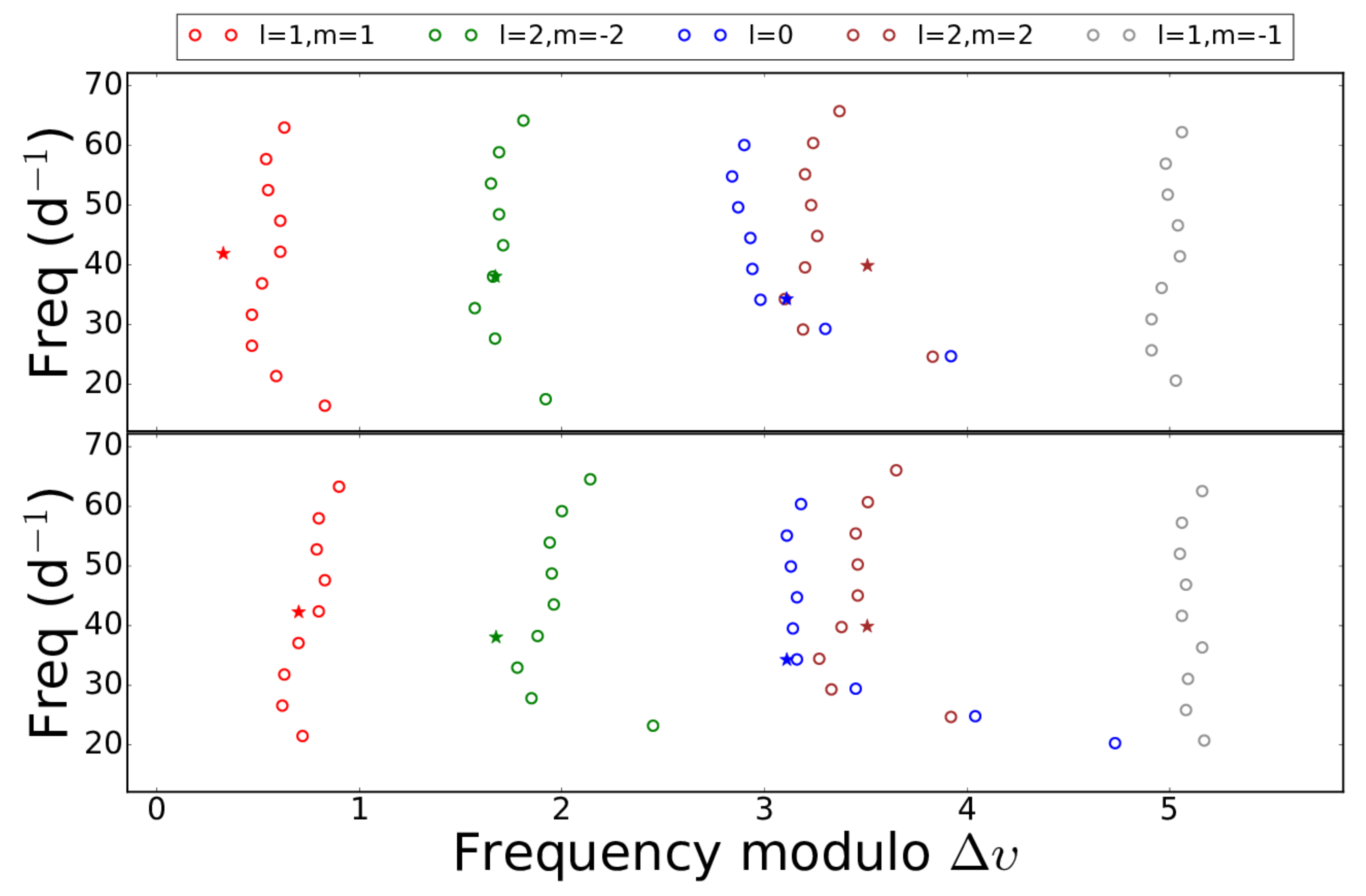}
    \caption{\red{Echelle diagram comparing scaled, calculated frequencies for Model B (upper panel) and the unscaled frequencies for Model D (lower panel). A large separation of frequency modulo\,=\,5.2 d$^{-1}$ is used in both cases.  The matching observed frequencies for TT Hor are indicated by the coloured stars with brown = \textit{f}$_1$; green = \textit{f}$_2$; blue =\textit{f}$_3$; red = \textit{f}$_4$. For each plot, The highest radial order shown is  \textit{n} =  10.} Positive $m$ values are prograde modes travelling in the direction of rotation.}

   \label{Echelle.eps}
 \end{figure}

 \begin{table*} 
	\centering
	\caption{Identification of the modes of the observed frequencies in the accretor by comparison to modes identified by \textsc{gyre} \red{after applying a scaling factor of either 2.7\% (Model B) or no scaling factor (Model D).} } 
	
	\label{tab:modeID}
	\begin{tabular}{ccccccccc} % 9 columns, alignment for each
		\hline
		
       & \red{Observed}  & Observed  & \multicolumn{2}{c}{Model B} && \multicolumn{2}{c}{\red{Model D}}  \\

       & Frequency & Amplitude  &\multicolumn{2}{c}{\textsc{gyre} Freq}  &&\multicolumn{2}{c}{\textsc{gyre} Freq}& Mode ID\\
        &(d$^{-1}$) & V (mmag) & Accretor & Single star & & Accretor & Single star\\
		\hline
		\textit{f}$_1$ & 39.907 $\pm$ 0.003 & 3.8 $\pm$ 0.3  & 39.596 	& 39.684 &   & 39.779&  39.760& \textit{n} = 5, \textit{l} = 2, \textit{m} = 2\\
		\textit{f}$_2$ &38.075 $\pm$ 0.003 & 2.8 $\pm$ 0.5 & 38.058		& 38.034 &	 & 38.282 & 38.262& \textit{n} = 5, \textit{l} = 2, \textit{m} = -2\\
		\textit{f}$_3$ &34.310 $\pm$ 0.003 & 2.2 $\pm$ 0.6 & 34.302  & 34.396 &  & 34.467& 34.450&\textit{n} = 4, \textit{l} = 2, \textit{m} = 2\\

     &&& 34.183 &	34.272 & & 34.362& 34.352& \textit{n} = 5, \textit{l} = 0\\
       \textit{f}$_4$ &42.301 $\pm$ 0.003 & 2.0 $\pm$ 0.1 & 42.207 	&42.248   &   & 42.404 & 42.423& \textit{n} = 6, \textit{l} = 1, \textit{m} = 1\\

		\hline
	\end{tabular}
\end{table*}

The observed frequencies for TT\,Hor are identified as  p\:modes,  tentatively order \textit{n} = 4, 5 or 6. (refer to Table~\ref{tab:modeID}).  All observed frequencies agree to within 0.7\% of the calculated ones.  \red{For Model B,} two frequencies, \textit{f}$_1$ and \textit{f}$_3$, are possibly the non-radial mode, \textit{l}\,=\,2,\textit{ m} = 2, with radial order \textit{n} = 5 or 4, respectively. 
However,  \textit{f}$_3$ = 34.309 d$^{-1}$ could also be an \textit{l}= 0  radial mode as the match is within 0.4\% of the calculated frequency. \red{These two options are clearly discriminated in Model D, with zero scaling, with \textit{f}$_3$ being within 0.15$\%$ of the \textit{n} = 5, \textit{l} = 0 radial mode and \textit{f}$_1$ within 0.3$\%$ of the \textit{n} = 5, \textit{l} = 2, \textit{m} = 2 mode.  } The other two frequencies \red{in both models} are non-radial modes. The sampling of the light curve was not conducive to the detection of g\:modes in this system.

\red{ Continuous, photometric time series over many days is expected to reveal many more frequencies and and we hope to pursue this with TESS or additional LCO data.  Such additional frequencies should enable more definitive mode identification and discriminate between adiabatic and non-adiabatic perturbation effects.  These, in turn, may lead to a more accurate binary model as we have shown how small changes in metallicity and mass can affect mode identification.}

As the rotational and orbital axes are aligned in the tidally locked system, then the pulsation axis will also be aligned.  The visibility of pulsations with different azimuthal orders (\textit{m}) is highly dependent on the inclination of the star to our line of sight.  \citet{gizon_determining_2003} showed that at high inclination (80\textdegree), the rotationally split modes, \textit{l} = 1, \textit{m} = $\pm$ 1 and \textit{l} = 2,\textit{ m}\,=\,$\pm$ 2, are the most likely observed in slowly rotating stars with angular velocities between 1 and 10 solar units ($\approx2$ - 20 km\,s$^{-1}$).  A sufficiently high rotation rate is needed to resolve azimuthal modes split by rotation, a condition met by TT Hor which is rotating at 41 km\,s$^{-1}$. However, not all modes are necessarily excited and therefore visible to the observer. Additionally, our data do not give reliable amplitudes for each frequency extracted via Fourier analysis, therefore substantiating our mode identification from their amplitudes is not plausible.  

The best matched frequency, \textit{f}$_2$, \red{for Model B}  is identified as an \textit{l} = 2, \textit{m} = -2 mode which has the second highest amplitude in our observations.   The lowest amplitude frequency is \textit{f}$_4$,  \textit{l} = 1, \textit{m} = 1. These preliminary mode identifications are in contrast to the findings of \citet{gizon_determining_2003} indicating that the \textit{l} = 1, \textit{m} = $\pm$ 1 should have a slightly higher amplitude than the \textit{l} = 2, \textit{m} = $\pm$\,2 modes at an inclination of 80\textdegree. Corroboration  of    \textit{f}$_4$ as an \textit{l} = 1, \textit{m} = 1 mode is not possible with our low pulsation amplitudes.  The frequency of 37.7 ($\pm$ 0.8)\,d$^{-1}$, identified during primary eclipse, showed an indication of an amplitude increase that could be associated with a spatial filtering effect. This frequency is closest to \textit{f}$_2$ = 38.071\,d$^{-1}$, which would be consistent with its identification as an \textit{l} = 2, \textit{m} = $-$2 mode. However, this effect needs to be confirmed.

In a circular orbit, the tidal effect is at equilibrium and not expected to excite g\:modes at high harmonics which are often found in systems with highly eccentric orbits (the so-called heartbeat stars, e.g. \citealt{hambleton_kic_2017}).    Such tidal effects have also been found in non-heartbeat stars such as CoRoT 105906206 \citep{da_silva_corot_2014} and KIC 9851944   \citep{guo_kepler_2016}, both of which have circularised orbits.  Tidally excited oscillations are most visible in stars with shallow surface convective zones when resonance-locking increases their amplitudes \citep{fuller_accelerated_2017}.  Resonance-locking may occur in circularised but non-synchronised binaries.  We found no orbital harmonics in the TT\,Hor data.

\section{Discussion and Conclusions}

We have determined a model of the semi-detached, eclipsing binary TT\,Hor determined from our photometry and RV measurements.  TT\,Hor consists of a 2.22 $M_{\sun}$ primary and a 0.72 $M_{\sun}$ secondary star and the high probability of a third body in distant orbit around the binary.  We confirm that the primary star (accretor) of the binary is the $\delta$ Sct pulsator and is an A4IV star.  

\red{Both our evolutionary models B and D match the binary model within uncertainties.  However, mode identification using Model B is only possible using a small density scaling factor, whereas for Model D, no scaling factor was needed to identify the same modes. Model D is therefore our preferred model for TT Hor, indicating that} the accretor and donor have evolved from an initial mass of \red{$\sim$1.29\,$M_{\sun}$ and $\sim$1.55 $M_{\sun}$, respectively.}  

The donor has lost \red{0.866} $M_{\sun}$ of slightly He-enriched material to the accreting, pulsating component which is evolving up the main sequence as it continues accreting mass from its companion.  TT\,Hor fits within the definition of \citet{mkrtichian_pulsating_2002, mkrtichian_frequency_2004} of an `oEA star', indicating pulsating, mass-accreting, main sequence stars of spectral type B-F in semi-detached Algol systems. 

The evolution of the $\delta$ Sct pulsator in TT\,Hor is quite different from that of a single, isolated $\delta$ Sct star, as shown by our evolutionary models.  Despite these differences, we are able to match the fundamental properties of the TT\,Hor accretor to a single $\delta$ Sct differing primarily only in age (0.30 and 2.58\,Gyr, respectively).  We are able to make a tentative identification of the pulsation modes of our observed frequencies using a MESA evolutionary model that matches our binary model.  TT Hor is a tidally-locked binary system, enabling an accurate determination of the rotation rate of the pulsator which is then applied to the mode identification.  Rotation rates cannot be measured to equivalent certainty in isolated stars and this is one of the major impediments in mode identification and modelling for $\delta$ Sct stars.  By applying the same rotation rate to the single star model, we are also able to associate our observed frequencies to the same pulsation modes.

The temperature of TT\,Hor ($T_{\rm eff}$ = 8800 $\pm$ 200 K) locates it at the extreme of the blue edge of the instability strip, $T_{\rm eff}$\,$\approx$\,8500\,K.  Here the HeII ionisation zone is close to the surface at $T_{\rm eff}$\,$\approx$\, 40,000\,K where the $\kappa$ mechanism is the driving mechanism and non-radial p\:modes are excited.  Radial orders of n = 5 or 6 are typically found \citep{xiong_turbulent_2016}, as is the case in our analysis.  We tentatively identify dipole and quadrupole non-radial modes, with the possibility of a radial mode.

Using \textit{Kepler} data, near-uniform surface-to-core rotation rates have been found in hybrid $\delta$~Sct and $\gamma$~Dor pulsators during their main-sequence evolution  \citep{kurtz_asteroseismic_2014,saio_asteroseismic_2015,murphy_near-uniform_2016}.   The stars in these studies have slow rotation periods, 27--100\,d making it easy to identify the rotational splitting of the modes. Once TESS data are available, we hope to identify g\:modes which would indicate any rotational differences in the core of TT\,Hor pulsator.  However, the pulsator is rotating moderately fast (41\,km\,s$^{-1}$) and disentangling the frequencies in the oscillation spectrum to identify rotational splitting may not be possible. 

Despite mass accretion and age differences between the TT\,Hor accretor and an isolated star, we have made a preliminary identification of the modes.  While these identified modes are by no means definitive, they are surprisingly close to the observed frequencies. Hence we have illustrated our premise that pulsators in Algol systems are powerful tools to validate models for isolated $\delta$ Sct stars.  We have data for other Algol-type systems which have different orbital and pulsation periods and by comparing these systems we anticipate shedding further light on these enigmatic pulsators.

\clearpage

\section*{Acknowledgements}

This research is supported by an Australian Government Research Training Program (RTP) Scholarship. SJM is the recipient of an Australian Research Council Discovery Early Career Award (project number 180101104), funded by the Australian Government.

This research was made possible through the use of the AAVSO Photometric All-Sky Survey (APASS), funded by the Robert Martin Ayers Sciences Fund.  
The Digitized Sky Surveys were produced at the Space Telescope Science Institute under U.S. Government grant NAG W-2166. The images of these surveys are based on photographic data obtained using the Oschin Schmidt Telescope on Palomar Mountain and the UK Schmidt Telescope. The plates were processed into the present compressed digital form with the permission of these institutions.
This publication makes use of data products from the Two Micron All Sky Survey, which is a joint project of the University of Massachusetts and the Infrared Processing and Analysis Center/California Institute of Technology, funded by the National Aeronautics and Space Administration and the National Science Foundation.
\red{This work has made use of data from the European Space Agency (ESA) mission
{\it Gaia} (\url{https://www.cosmos.esa.int/gaia}), processed by the {\it Gaia}
Data Processing and Analysis Consortium (DPAC,
\url{https://www.cosmos.esa.int/web/gaia/dpac/consortium}). Funding for the DPAC
has been provided by national institutions, in particular the institutions
participating in the {\it Gaia} Multilateral Agreement.}
%%%%%%%%%%%%%%%%%%%%%%%%%%%%%%%%%%%%%%%%%%%%%%%%%%

%%%%%%%%%%%%%%%%%%%% REFERENCES %%%%%%%%%%%%%%%%%%

% The best way to enter references is to use BibTeX:

\bibliographystyle{mnras}
\bibliography{TT_Hor_references}{}

\appendix

\section{Radial Velocity measurements}
 \begin{table} 
	\centering
	\caption{Radial velocities as determined with \textsc{todcor}. RV1 and RV2 are radial velocities for the primary and secondary stars, respectively. }  
	
	\label{tab:refRV}
	\begin{tabular}{ccrr} % four columns, alignment for each
		\hline
       Date (HJD)&	Orbital phase	& RV1	& RV2 \\
       \hline
      2456968.95762	&0.10	&5.8	&145.6\\
2456969.00862	&0.12	&-2.4	&145.2\\
2456969.19061&	0.19&	-26.5&	169.4\\
2456969.21061&	0.20&	-25.0&	181.0\\
2456969.89460&	0.46&	13.1&	63.3\\
2456969.90860&	0.47&	11.5&	65.8\\
2456969.92660&	0.47&	14.3&	68.9\\
2456969.94460&	0.48&	14.1&	64.7\\
2456970.91559&	0.85&	72.3&	-106.4\\
2456970.94259&	0.86&	72.9&	-92.0\\
2456970.96959&	0.87&	70.3&	-79.3\\
2456970.98758&	0.88&	67.2&	-68.4\\
2456971.23258&	0.98&	48.4&	15.1\\
2456971.89957&	0.23&	-28.2&	189.2\\
2456971.92357&	0.24&	-29.7&	185.9\\
2456971.94557&	0.25&	-28.6&	190.8\\
2456971.96057&	0.25&	-28.3&	196.4\\
2456971.99057&	0.27&	-25.5&	201.4\\
2456972.18056&	0.34&	-18.9&	172.3\\
2456972.21456&	0.35&	-17.8&	166.0\\
2456972.22756&	0.36&	-14.8&	166.5\\
2456972.23356&	0.36&	-14.9&	166.1\\
2456972.24656&	0.36&	-14.7&	167.3\\
2456975.98349&	0.80&	86.3&	-129.5\\
2456976.15848&	0.87&	77.2&	-75.8\\
2456976.97547&	0.18&	-20.1&	168.1\\
2456977.91745&	0.54&	40.0&	19.9\\
2456978.93543&	0.93&	60.4&	-14.8\\
2456978.95542&	0.94&	56.9&	-7.2\\
2456978.97242&	0.94&	50.2&	-3.8\\
2456979.00542&	0.96&	48.5&	9.7\\
2456979.03242&	0.97&	46.3&	15.9\\
2456979.05742&	0.98&	43.2&	19.9\\
2456979.09042&	0.99&	39.0&	27.3\\
2456979.11642&	1.00&	28.6&	31.6\\
2456979.17342&	0.02&	18.5&	41.7\\
2456979.21142&	0.03&	8.9	&53.9\\
2456979.25241&	0.05&	6.6	&64.3\\
2457678.08207&	0.98&	27.5&	6.1\\
2457679.05127&	0.36&	-24.1&	159.6\\
2457679.06517&	0.36&	-28.67&	148.4\\
2457679.08207&	0.37&	-22.9&	146.1\\
2457679.12297&	0.38&	-19.4&	132.5\\
2457679.13897&	0.39&	-24.6&	121.6\\
2457679.15488&	0.40&	-18.8&	106.7\\
2457679.17107&	0.41&	-18.2&	99.0\\
2457679.19037&	0.41&	-15.3&	89.6\\
2457679.20407&	0.41&	-21.8&	70.8\\
2457679.22197&	0.42&	-15.6&	89.9\\
2457679.25297&	0.43&	-0.2&	60.1\\
2457680.09407&	0.76&	70.7&	-154.1\\
2457680.10897&	0.76&	72.9&	-147.3\\
2457680.12727&	0.77&	67.8&	-128.9\\
2457680.14747&	0.78&	65.5&	-154.2\\
2457680.16587&	0.78&	66.7&	-140.9\\
2457680.18897&	0.79&	68.8&	-138.0\\
2457680.21047&	0.80&	66.5&	-140.6\\
2457680.22417&	0.81&	65.5&	-136.4\\
2457680.24567&	0.81&	68.2&	-126.8\\
2457680.25987&	0.82&	67.1&	-128.9\\
 
	\hline
    
	\end{tabular}
\end{table}
%
%If you want to present additional material which would interrupt the flow of the main paper, it can be placed in an Appendix which appears after the list of references.

%%%%%%%%%%%%%%%%%%%%%%%%%%%%%%%%%%%%%%%%%%%%%%%%%%

% Don't change these lines
\bsp	% typesetting comment
\label{lastpage}
\end{document}